\documentclass[12pt]{article}
\begin{document}
\title{Modelling Time}
\author{B.G. Sidharth\\
International Institute for Applicable Mathematics \& Information Sciences\\
Hyderabad (India) \& Udine (Italy)\\
B.M. Birla Science Centre, Adarsh Nagar, Hyderabad - 500 063
(India)}
\date{}
\maketitle
\begin{abstract}
We briefly review two concepts of time - the usual time associated
with "being" and more recent ideas, answering to the description of
"becoming". The approximation involved in the former is examined.
Finally we argue that it is (unpredictable) fluctuations that
underlie time.
\end{abstract}
\vspace{5 mm}
\begin{flushleft}
\end{flushleft}
\section{Introduction}
Twentieth century physics inherited the Newtonian concept of space
and time. Even with the advent of Special Relativity some of these
ideas remained though time became less absolute and the concept of
simultaneity got modified. In particular time was reversible. With
the advent of Quantum Mechanics, there was a carry over of these
concepts, even though there were new inputs like the effect of
observation. So the description is that of a time reversible
universe, in contrast to real life experience. It may be mentioned
however that all this is strictly speaking valid for a single
particle universe or at best an isolated system of a few particles.
In statistical mechanics we encounter a different scenario where
time is irreversible. Indeed the arrow of time is
associated with an increase of entropy.\\
All this can be brought out by the simple example of a container
with two compartments. Compartment $A$ is filled with a swarm of
molecules. In contrast compartment $B$ is empty. A tiny hole is then
bored into the dividing wall. Molecules from $A$ gradually leak into
compartment $B$, till both compartments have more or less an equal
number of molecules. Molecules would still be crossing over from $A$
to $B$ and vice versa, but this state of equilibrium persists. Here
with the passage of time the empty compartment $B$ has got filled up
with molecules. There is however no reversibility as in classical
and Quantum Theory in the sense that we do not get back a situation,
even in principle where the compartment $B$ gets emptied out and all
the molecules in $B$ fill up compartment $A$.\\
There have been a number of arguments for and against the arrow of
time in the context of modern theories of physics
\cite{bgstimes,hugh,tiemblo,land,penrose}. Nevertheless the fact
that there should not be an arrow of time even within the context of
Quantum Theory has been discussed \cite{bgstimes}. Indeed as Penrose
prophesized, \cite{hawking} "the tiny fact of an almost completely
hidden time-asymmetry seems genuinely to be present in the
$K^0$-decay. It is hard to believe that nature is not, so to speak,
trying to tell something through the results of this delicate and
beautiful experiment." It may be observed that the time
symmetries actually encountered are:\\
i) The thermodynamic arrow of time, which represents an evolution
towards a greater entropy.\\
ii) The electromagnetic arrow of time, which becomes apparent in the
retarded nature of electromagnetic radiation chosen due to considerations of
causality.\\
iii) The CP violating arrow of time.\\
iv) The cosmological arrow of time which appears in the expansion of
the universe or the formation of structures.\\
These arrows are generally sought to be explained in terms of the
usual time symmetric dynamics, but with the additional input of time
asymmetric boundary conditions, for instance in the above example of
a container $A$ with molecules being in contact with an empty
container $B$, we start ab initio with an asymmetric situation. The
initial asymmetry in the universe is provided by the Big Bang, as
another example \cite{fink1,fink2}. This is also the content of the
Ehrenfest's so called double urn model \cite{fink3}--it is
essentially the above two compartment argument. In any case this is
how an asymmetric outcome is understood to arise from symmetric
considerations--the arrow of time from time reversible laws. Other
models have not lead to anything conclusive, for example with the
rather abstract arguments of Finkelstein \cite{fink4}.
\section{Prigogine's Analysis}
Prigogine had analyzed the above situation quite clearly. He
observed \cite{prig} that according to the classical point of view,
nature would be an automaton. However, today we discover everywhere
instabilities, bifurcations, evolution. This demands a different
formulation of the laws of nature to include probability and time
symmetry breaking. He argued that the difficulties in the classical
formulation come from too narrow a point of view concerning the
fundamental laws of dynamics (classical or quantum). The classical
model has been a model of integrable systems (in the sense of
Poincare). It is this circumstance that leads to determinism and
time reversibility. He argued that when we leave this model and
consider a class of non-integrable systems, the difficulties are
overcome. Further, this approach unifies dynamics, thermodynamics
and probability theory.\\
The point is that in classical theory we use unitary transformations
(Cf.ref.\cite{prig,gold}) to introduce new momenta and coordinates
related to the old $p$ and $q$ by $p' = U^{-1} p, q' = U^{-1} q$,
where $U$ is a so-called unitary operator. These transformations are
made in such a way that the Hamiltonian equations remain valid. $U$
plays an essential role both in classical and quantum mechanics. An
important property is the distributivity of $U$. That means that $U$
acting on a product is equal to the product of the individual
transformations. That is, $U^{-1} (AB) = (U^{-1}A)(U^{-1}B)$. The
above formulation of classical mechanics and integrable systems is
taken over into Quantum Mechanics though we no longer use numbers
$p$ and $q$ but rather operators. The important point however is
that
Quantum Mechanics too uses integrable systems.\\
Prigogine goes on to point out that integrable systems are the
exception, rather than being the rule. Let us consider a simple
example of a non integrable system viz., a Harmonic oscillator in
for example an electromagnetic field. If the frequency of the
Harmonic oscillator $\omega$ is outside of the spectrum of the
frequencies $\omega'$ of the field then there is no new situation.
When however the oscillator frequency falls within the frequency
spectrum of the field, then we have to deal with divergent terms
like $1(/\omega - \omega')$. This leads to the well known Poincare
resonances.\\
We could still continue with a transformation like the unitary
transformation $U$, except that it gets replaced by a transformation
$\Lambda$ which is not unitary but rather is called star unitary.
There is now an extension of canonical transformations and the old
theory no longer applies. The operator $\Lambda$ becomes non local.
As Prigogine observes, "...classically people were thinking in terms
of points but here we have to speak in terms of ensembles. We cannot
any more make a physics of points but we have to make a physics of
distributions. This means that we have a statistical description.
That also means that we have to give up classical determinism."
Other new elements
are also introduced into the theory.\\
It must be mentioned that even in Quantum Theory, be it the
Schrodinger equation or the Dirac equation, the wave function still
follows a deterministic theory as in classical mechanics. The
indeterminism of Quantum Mechanics comes in through a different
mechanism, that is an observation disturbs the system and there is
at that very instant a non causal collapse of the wave function into
one of its eigen states. Interestingly it has been shown
\cite{kim,ordonez} that from the new non unitary transformations one
can derive the Langevin equations and this leads to a contact with
statistical mechanics. It is also possible to obtain the Quantum
Langevin equation.
\section{Microscopic Considerations}
In the preceding section, the analysis was based on notions of
macroscopic mechanics. However Salecker and Wigner
\cite{wigner,salecker} argued that the macroscopic concepts of space
and time have no operational meaning for micro systems. That is
spacetime descriptions may be valid only for macroscopic systems. To
quote Wigner, "... the inherent limitations on the accuracy of a
clock of given weight and size, which should run for a period of a
certain length, are quite severe. In fact, the result in summary is
that a clock is an essentially non-microscopic object. In
particular, what we vaguely call an atomic clock, a single atom
which ticks off its periods, is surely an idealization which is in
conflict with fundamental concepts of measurability."\\
Zimmerman and others \cite{zimmer,karimov} suggested that space and
time arise from the properties of microscopic particles in a
thermodynamic fashion in the sense that the former are a result of
interactions among many microscopic systems, without a detailed
description of the interactions amongst these systems in
spatio-temporal terms.\\
In this context there have been more recent efforts in Quantum
Gravity approaches. While some resemble Hawking's Euclidean Quantum
Gravity in that the super position principle is applied for the
several possible ways the entire universe could evolve in time
(Cf.refs.\cite{halliwell} for a readable account), a more recent
variant approach has been that of causal dynamical triangulation of
Loll, Ambjorn and others. Here in the spirit of the usual
triangulation, spacetime develops from the micro level in a self
organizational scheme \cite{adrien,ambjorn,loll}. It must be
mentioned that these recent approaches are based on three pillars
which are, discreteness rather than continuum, emergence (as opposed
to a reductionist approach) and causality. While this is reasonable
from a physical point of view, particularly the requirement of
causality does not go far enough. It already hides an implicit arrow
of time. In any case these approaches cannot as yet be termed as
being successful or the last word \cite{smolin}.\\
 In this spirit the author has argued
(Cf.\cite{tduniv} and several references therein) that physical
concepts of space and time or spacetime arise outside the Compton
wavelength within which these concepts no longer hold
\cite{bgstlsr}. Indeed this is already implicit in Wigner's analysis
referred to. Within the Compton wavelength, we have unphysical
phenomena like Zitterbewegung and a breakdown of causality- only an
average over the Compton scale restores physics (Cf. also
\cite{tduniv,revzitter}). Within the Compton scale time can be
modelled by a double Weiner process corresponding to a one
dimensional Random Walk. It must be reiterated that within the
Compton wavelength there is no causal physics \cite{weinberggrqc}.
Indeed it can be argued that Special Relativity becomes operational
outside the Compton wavelength (Cf.ref.\cite{bgstlsr}). In this
sense we are justified in
considering time, rather than spacetime. We will return to this shortly.\\
To appreciate all this let us consider the motion of a particle with
position given by $x(t)$, subject to random correction given by, as
in the usual theory, (Cf.\cite{smolin2,reif,heap}),
$$|\Delta x| = \sqrt{<\Delta x^2 >} \approx \nu \sqrt{\Delta t},$$
\begin{equation}
\nu = \hbar/m, \nu \approx l v\label{2e3}
\end{equation}
where $\nu$ is the so called diffusion constant and is related to
the mean free path $l$ as above. We can then proceed to deduce the
Fokker-Planck equation as follows (Cf.ref.\cite{smolin2} for
details):\\
We first define the forward and backward velocities corresponding to
having time going forward and backward (or positive or negative time
increments) in the usual manner,
\begin{equation}
\frac{d_+}{dt} x (t) = {\bf b_+} \, , \, \frac{d_-}{dt} x(t) = {\bf
b_-}\label{2ex1}
\end{equation}
This leads to the Fokker-Planck equations
$$
\partial \rho / \partial t + div (\rho {\bf b_+}) = V \Delta \rho
,$$
\begin{equation}
\partial \rho / \partial t + div (\rho {\bf b_-}) = - U \Delta
\rho\label{2ex2}
\end{equation}
defining
\begin{equation}
V = \frac{{\bf b_+ + b_-}}{2} \quad ; U = \frac{{\bf b_+ - b_-}}{2}
\label{2ex3}
\end{equation}
We get on addition and subtraction of the equations in (\ref{2ex2})
the equations
\begin{equation}
\partial \rho / \partial t + div (\rho V) = 0\label{2ex4}
\end{equation}
\begin{equation}
U = \nu \nabla ln\rho\label{2ex5}
\end{equation}
It must be mentioned that $V$ and $U$ are the statistical averages
of the respective velocities. We can then introduce the definitions
\begin{equation}
V = 2 \nu \nabla S\label{2ex6}
\end{equation}
\begin{equation}
V - \imath U = -2 \imath \nu \nabla (l n \psi)\label{2ex7}
\end{equation}
We next observe the decomposition of the Schrodinger wave function
as
$$\psi = \sqrt{\rho} e^{\imath S/\hbar}$$
leads to the well known Hamilton-Jacobi type equation
\begin{equation}
\frac{\partial S}{\partial t} = -\frac{1}{2m} (\partial S)^2 + {\bf
\bar{V}}+Q,\label{2e4}
\end{equation}
where
$$Q = \frac{\hbar^2}{2m} \frac{\nabla^2 \sqrt{\rho}}{\sqrt{\rho}}$$
From (\ref{2ex6}) and (\ref{2ex7}) we can finally deduce the usual
Schrodinger equation or (\ref{2e4}) \cite{nottalefractal}.\\
We note that in this formulation three conditions are assumed,
conditions whose import has not been clear. These are \cite{smolin2}:\\
(1) The current velocity is irrotational. Thus, there exists a
function $S(x,t)$ such that
$$m \vec V = \vec \nabla S$$
(2) In spite of the fact that the particle is subject to random
alterations in its motion there exists a conserved energy, defined
in terms of its
probability distribution.\\
(3) The diffusion constant is inversely proportional to the inertial
mass of the particle, with the constant of proportionality being a
universal constant $\hbar$ (Cf. equation (\ref{2e3})):
$$\nu = \frac{\hbar}{m}$$
We note that the complex feature above disappears if the fractal or
non-differentiable character is not present, (that is, the forward
and backward time derivatives(\ref{2ex3}) are equal): What
distinguishes Quantum Mechanics is the adhoc feature, the diffusion
constant $\nu$ of (\ref{2e3}) in Nelson's theory and the "Quantum
potential" $Q$ of (\ref{2e4}) which appears in Bohm's theory
\cite{cu} as well, though with a
different meaning.\\
Interestingly from the Uncertainty Principle,
$$m \Delta x  \frac{\Delta x}{\Delta t} \sim \hbar$$
we get back equation (\ref{2e3}) of Brownian motion. This shows the
close connection on the one hand, and provides, on the other hand, a
rationale for the particular, otherwise adhoc identification of
$\nu$ in (\ref{2e3}) - its being proportional to $\hbar$.\\
We would like to emphasize that we have arrived at the Quantum
Mechanical Schrodinger equation from Classical considerations of
diffusion, though with some new assumptions. In the above,
effectively we have introduced a complex velocity $V - \imath U$
which alternatively means that the real coordinate $x$ goes into a
complex coordinate
\begin{equation} x \to x + \imath x'\label{2De9d}
\end{equation}
To see this in detail, let us rewrite (\ref{2ex3}) as
\begin{equation}
\frac{dX_r}{dt} = V, \quad \frac{dX_\imath}{dt} = U,\label{2De10d}
\end{equation}
where we have introduced a complex coordinate $X$ with real and
imaginary parts $X_r$ and $X_\imath$, while at the same time using
derivatives with respect
to time as in conventional theory.\\
We can now see from (\ref{2ex3}) and (\ref{2De10d}) that
\begin{equation}
W = \frac{d}{dt} (X_r - \imath X_\imath )\label{2De11d}
\end{equation}
That is, in this non relativistic development either we use forward
and backward time derivatives and the usual space coordinate as in
(\ref{2ex3}), or we use the derivative with respect to the usual
time coordinate but introduce complex space coordinates as in
(\ref{2De9d}). Already, we can get a glimpse of the special
relativistic hyperbolic geometry with real space and imaginary time
coordinates
(or vice versa).\\
We now try to generalize this \index{complex coordinate}complex
coordinate to three dimensions. Then we encounter a surprise - we
end up with not three, but four dimensions,
$$(1, \imath) \to (I, \sigma),$$
where $I$ is the unit $2 \times 2$ matrix and $\sigma$s are the
Pauli matrices. We get the special relativistic
\index{Lorentz}Lorentz invariant metric at the same time. (In this
sense, as noted by Sachs \cite{sachsgr}, Hamilton who made this
generalization would have hit upon \index{Special Relativity}Special
Relativity, if he had identified the new fourth coordinate
with time).\\
That is,\\
$$x + \imath y \to Ix_1 + \imath x_2 + jx_3 + kx_4,$$
where $(\imath ,j,k)$ now represent the \index{Pauli}Pauli matrices;
and, further,
$$x^2_1 + x^2_2 + x^2_3 - x^2_4$$
is invariant. Before proceeding further, we remark that special
relativistic time emerges above from the generalization of the
complex one dimensional space coordinate to three dimensions.\\
While the usual \index{Minkowski}Minkowski four vector transforms as
the basis of the four dimensional representation of the
\index{Poincare}Poincare group, the two dimensional representation
of the same group, given by the right hand side in terms of
\index{Pauli}Pauli matrices, obeys the quaternionic algebra of the
second rank
\index{spin}spinors (Cf.Ref.\cite{bgsfpl162003,shirokov,sachsgr} for details).\\
In the above context let us try to see what the time of the usual
theory is: In the stochastic approach, we deal with a double
\index{Wiener process}Wiener process which leads to a complex
velocity $V-\imath U$. As noted it is this complex velocity that
leads to \index{Quantum Theory}Quantum Theory from the usual
\index{diffusion}diffusion theory.\\
To see this in a simple way, let us write the usual
\index{diffusion}diffusion equation as
\begin{equation}
\Delta x \cdot \Delta x = \frac{h}{m}\Delta t \equiv \nu \Delta
t\label{5eb1}
\end{equation}
We saw that equation (\ref{5eb1}) can be rewritten as the usual
Quantum Mechanical relation,
\begin{equation}
m\frac{\Delta x}{\Delta t} \cdot \Delta x = h = \Delta p \cdot
\Delta x\label{5eb2}
\end{equation}
We are dealing here, with phenomena within the Compton or \index{De
Broglie}De Broglie
wavelength.\\
We now treat the \index{diffusion}diffusion constant $\nu$ to be
very small, but non vanishing. That is, we consider the semi
classical case. This is because,
a purely classical description, does not provide any insight.\\
It is well known that in this situation we can use the WKB
approximation \cite{schiff}. Whence the right hand side of the wave
function,
$$\psi = \sqrt{\rho} e^{\imath /\hbar S}$$
goes over to, in the one dimensional case, for simplicity,
$$(p_x) ^{-\frac{1}{2}} e^{\frac{1}{h}} \int p(x)dx$$
so that we have, on comparison,
\begin{equation}
\rho = \frac{1}{p_x}\label{5eb3}
\end{equation}
$\rho$ being the probability density. In this case the condition $U
\approx 0$, that is, the velocity potential becoming real, implies
\begin{equation}
\nu \cdot \nabla ln (\sqrt{\rho}) \approx 0\label{5eb4}
\end{equation}
This semi classical analysis suggests that $\sqrt{\rho}$ is a slowly
varying function of $x$, in fact each of the factors on the left
side of (\ref{5eb4}) would be $\sim 0(h)$, so that the left side is
$\sim 0(h^2)$ (which is being neglected). Then from (\ref{5eb3}) we
conclude that $p_x$ is independent of $x$, or is a slowly varying
function of $x$. The equation of continuity now gives
$$\frac{\partial \rho}{\partial t} + \vec \nabla (\rho \vec v)  = \frac{\partial \rho}
{\partial t} = 0$$ That is the probability density $\rho$ is
independent or nearly so, not only of $x$¬ but also of $t$. We are
thus in a stationary and homogenous scenario. This is strictly
speaking, possible only in a single particle
\index{Universe}Universe, or for a completely isolated particle,
without any effect of the environment. Under these circumstances we
have the various conservation laws and the time reversible theory,
all this taken over into \index{Quantum Mechanics}Quantum Mechanics
as well. This is an approximation valid for small, incremental
changes, as indeed is implicit in the concept of a
differentiable spacetime manifold.\\
To put it simply, if $dt$ is the change or time interval in the
usual time, then $N$ such intervals would imply a passage of time of
magnitude $N dt$, whereas in our approach, if $d \tau$ is the basic
interval, then the time passage would be $\sqrt{N} d \tau$.\\
We recognize that time is statistical and depends on the number of
constituents. There could be collisions amongst $N$ particles, or
equivalently, as we will see below the fluctuational creation of
such particles from the background Dark Energy. It could then be
possible to ascribe a general minimum time interval $\tau$ for all
the $N$ particles, which would be for example the mean free time
between the collisions or the fluctuational creation of particles.
On the other hand let $\tau_0$ represent the corresponding quantity,
but this time associated with each individual event rather than the
entire assembly. We would then have (Cf. also ref.\cite{karimov})
\begin{equation}
\tau = \lambda \tau_0,\label{ea}
\end{equation}
where $\lambda$ represents the statistical dispersion effect and is
given by
\begin{equation}
\lambda = \left[\frac{1}{N} \sum^{N}_{\imath = 1} (\theta_\imath)^2
- \left(\frac{1}{N}\sum^{N}_{\imath =
1}\theta_\imath\right)^2\right]^{1/2}\label{eb}
\end{equation}
$\theta_\imath$ representing the coefficient to be multiplied into
each individual $\tau_0$, in order to get the corresponding interval
for the $\imath$th particle. As $N$ is large and the $\theta_\imath$
are all of the order $1$, using (\ref{eb}) in (\ref{ea}) we get
\begin{equation}
\tau = 1/\sqrt{N} \tau_0\label{ec}
\end{equation}
As there are $N$ such events, the time elapsed $T$ would be $N \tau$
which from (\ref{ec}) is given by
\begin{equation}
T = \sqrt{N} \tau_0\label{ed}
\end{equation}
The relevance of all this is the following. The author's 1997 model
was one in which particles were fluctuationally created from a
background Dark Energy within the Compton time $\tau_0$. It may be
mentioned that this model lead to a cosmology in which the universe
was accelerating with a small cosmological constant, besides many
other consistent but otherwise inexplicable astrophysical relations
being deduced from the theory (Cf.refs.\cite{bgsijmpa,cu,uof,tduniv}
and several references therein). As is known all this was confirmed
in 1998 itself through the observations of distant supernovae. In
this model given $n$ particles at any time, $\sqrt{n}$ particles
would be fluctuationally created in the time $\tau_0$. That is we
would have,
$$\frac{dn}{dt} \approx \frac{\sqrt{n}}{\tau_0}$$
because the interval $\tau_0$ being small, we can approximate with
derivates. Whence on integration from time $t = 0$ to $t = T$, we
get,
\begin{equation}
T = \sqrt{N} \tau_0\label{ee}
\end{equation}
In (\ref{ee}), at time $T$ there would be $N$ particles in the
universe. Not only is (\ref{ee}) consistent, because there are $N$
of the order $10^{80}$ particles in the universe while $T$ is of the
order $10^{17}$ seconds and $\tau_0$ a typical Compton time is of
the order $10^{-23}$seconds, but as can be seen (\ref{ee}) is
identical to (\ref{ed}) though it is obtained by a different route.
The above is very much in the spirit of a one dimensional Random
Walk or two Weiner process encountered earlier. In this case time
would flip forward and backward like steps to the right and steps to
the left at random. The nett dispersion or time elapsed after $N$
such stages would be exactly as in equations (\ref{ed}) or
(\ref{ee}). In the above cosmological model this would mean that
particles would be created and destroyed at random, from and into
the background Dark Energy, but the nett result is given by
(\ref{ee}). In this formulation, there is no inbuilt causality.
\section{Remarks}
Time is associated with change. In a changeless universe, there
would be no time. It must be mentioned here that various theories of
time, already imply time or more generally change (Cf. also
\cite{aiello}). The question is, what type of a change do we
consider? We have argued that the time which we usually use is based
on an incremental change - it is almost as if there were no change.
For example the law of conservation of energy is based on a time
translation symmetry - an infinitesimal translation in time leaves
everything unchanged \cite{roman}. Clearly this is only an
approximation which assumes that there is no change in a very short
interval.\\
In fact time is essentially an ordering or sequencing of events. The
key here is the way in which this ordering is done so that causality
and other laws of physics hold or emerge rather than being inputs.
If we consider the universe as a sequence of instantaneous space
slices, to start with, then a random sequence would represent a
lawless, and literally chaotic universe.\\
On the contrary we have seen that at the micro scale, that is, more
specifically within the Compton scale, indeed there is no causality
and there is the chaotic Zitterbewegung. This means that if it were
possible for a creature or a measuring device to be so small as to
be within the Compton scale, then such a creature or device would
perceive a lawless, chaotic universe. However physics, (and this
includes elementary particles) emerges once averages over the
unphysical Compton scale are taken. In this sense the universe that
is perceived and measured is a macroscopic universe. At the micro
level, as pointed out by Wigner and Sackler, we can no longer
extrapolate these macro concepts. In a sense, this is connected with
the Copenhagen debate on the role of macroscopic measuring devices
in
obtaining information about microscopic systems.\\
At the macro scale however we have two different situations. One,
which we encountered in the first section can further be exemplified
with the example of a porcelain plate that falls to the ground and
breaks into many pieces. Here a highly ordered system, namely the
porcelain plate becomes a highly disordered system, namely the
collection of shattered pieces. As long as we do not specify the
exact shapes and sizes of the shattered pieces in advance, this can
always happen - it provides an arrow of time with increasing
entropy. This time, furthermore is irreversible. The shattered
pieces then combining to form the plate or equivalently the
shattered pieces, which describe a very definite prescribed shape
and size would be an impossibility and would represent the reversal
of time. In any case, the connection between time and probability is
brought about. If the probability of something happening is (in
advance) zero or nearly so, time does not "evolve" to such a
situation. Time's
arrow or flow is in the direction of non vanishing probabilities.\\
There is another change that we had considered in the previous
section - this is the fluctuational creation of particles. This
gives rise to a macroscopic time, through a Brownian process,
leading to the correct age of the universe as exemplified in
(\ref{ee}). In contrast to the change which our usual time
represents, this latter Brownian change is no longer incremental.
This picture is a far cry from the smoothly flowing time of usual
theory. The contrast is between "becoming" and "being". Moreover,
this time is based on, not just the local particle, but a whole
assembly of particles, as brought out by (\ref{eb}), for example.
Our contention is that it is this irreversible change of
fluctuations that represents our actual time. This is in the spirit of
Bergson's future being creation \cite{prigogine}.\\
In any case, the above discussion shows the breakdown of the concept
of point time or instants of time. Indeed as Wigner (loc.cit) notes,
this concept has no meaning at least within the time interval thrown
up by the Uncertainty Principle, and perhaps even beyond. Finally,
it may be mentioned that the probabilistic feature of time discussed
above, brings in to play via observer participation, two poorly
understood concepts as of now: information and consciousness.
\vspace{5 mm}
\begin{flushleft}
{\large {\bf ACKNOWLEDGEMENT}}
\end{flushleft}
\noindent I am grateful to the editor for the pertinent and valuable
comments in the light of which the paper has been revised.

\end{document}